\documentstyle[12pt,epsf]{article}
\newcommand{\simgeq}{\; \raisebox{-0.4ex}{\tiny$\stackrel
{{\textstyle>}}{\sim}$}\;}
\newcommand{\simleq}{\; \raisebox{-0.4ex}{\tiny$\stackrel
{{\textstyle<}}{\sim}$}\;}
\newcommand{\beq}{\begin{equation}}
\newcommand{\beqar}{\begin{eqnarray}}
\newcommand{\eeq}[1]{\label{#1} \end{equation}}
\newcommand{\eeqar}[1]{\label{#1} \end{eqnarray}}
\begin{document}
\baselineskip 25pt minus .1pt
\begin{center}
{\large{\bf Kinetic energy and spin-orbit splitting in 
nuclei near neutron drip line}} \\
[4ex]

{\bf I.Hamamoto$^a$, S.V.Lukyanov$^{a,b}$ and X.Z.Zhang$^{a,c}$} \\ [2ex]

$^a$ Department of Mathematical Physics \\ Lund Institute of Technology at 
University of Lund \\ Lund, Sweden. \\ [2ex]

$^b$ Institute for Nuclear Research of the National Academy of Sciences\\ 
Kiev, Ukraine

$^c$ Institute of Atomic Energy, Beijing \\ The People's Republic of China.

\end{center}

\vspace{2cm}

\noindent
ABSTRACT : 
Two important ingredients of nuclear shell-structure, 
kinetic energy and spin-orbit splitting, are studied 
as a function of orbital angular momenta $\ell$ and binding 
energies, when binding energies of neutrons 
decrease towards zero.
If we use the standard parameters of the Woods-Saxon 
potential in $\beta$ stable nuclei  
and approach the limit of zero binding energy from 10 MeV, 
the spin-orbit splitting for $n$=1 orbitals decreases considerably for   
$\ell$=1, while for $\ell$$>$2 little decreasing is observed in 
the limit.  In contrast, 
the kinetic energy decreases considerably for $\ell$$\simleq$3.  
The smaller the $\ell$ values of orbitals, the larger the decreasing rate 
of both kinetic energy and spin-orbit splitting.
The dependence of the above observation on the diffuseness of potentials is
studied.

\vspace{1cm}

\noindent
PACS numbers~ : ~21.10.Pc

\newpage
\section{Introduction}
\indent
One-particle shell structure is well studied for nuclei along the $\beta$
stability line.
Using recent 
radioactive nuclear ion beam facilities it has become possible to study 
the structure of exotic nuclei away from the $\beta$ stability line, in which
least-bound 
nucleons have much smaller separation energies compared with those in $\beta$
stable nuclei.
Some interesting change of shell structure is expected and/or observed, when
we move away from $\beta$ stable nuclei to drip line nuclei.  This change is 
made due to the
presence of loosely-bound nucleons 
as well as exotic values of isospin for a given
mass number.
For example, available experimental information 
on the nucleus $^{32}_{12}$Mg$_{20}$, 
which is a singly-closed-shell nucleus in the traditional terminology, 
seems to indicate that the nucleus is deformed \cite{GM84,MOT95}.
Furthermore, the creation of a new magic number N=16 
near the neutron drip line (T$_{z}$$\geq$3) is recently suggested 
\cite{Ozawa00}.
In the present work we study the two basic ingredients 
of nuclear shell-structure,
kinetic energy and spin-orbit splitting, for relatively simple
potentials, having the spirit similar to that of ref. \cite{CDV96}.
In available publications certain aspects of 
shell structure unique in drip line nuclei have been studied 
using self-consistent Hartree-Fock (HF) calculations of various types.

\indent
Using Skyrme-type HF calculations, in ref. \cite{HSZ96} 
we have studied neutron one-particle level scheme
with a fixed mass number as a function of neutron number, including
one-particle resonant levels.
It is seen that calculated 
one-particle levels with lower 
$\ell$ are systematically pushed down relative to those with higher $\ell$, as
we approach the neutron drip line. 
On the other hand, in 
ref. \cite{LVP98} the reduction of the spin-orbit potential in light neutron
rich nuclei, especially the isospin-dependence of the spin-orbit term is
discussed, using the relativistic mean field theory.  
In the Skyrme-type 
HF calculations of ref. \cite{HS99} it is found that the estimated 
1p$_{1/2}$-1p$_{3/2}$ spin-orbit splitting  of the light mirror nuclei,
$^8_6$C$_2$ and $^8_2$He$_6$, is about 30 percent smaller than the one in the
$\beta$ stable nucleus $^{12}_6$C$_6$.
The difference of the isospin-dependence of spin-orbit term in the relativistic
and traditional Skyrme-type HF calculations is certainly important \cite{RF95} 
to be examined. 

\indent
A systematic study of both kinetic energy and spin-orbit splitting 
as a function of one-particle binding energy, orbital angular momentum 
and the shape of
the potential is essential in understanding 
the variation of the shell structure
as we approach the drip lines. 
In self-consistent HF calculations not only the orbitals with less
binding energies are occupied but also the effective diffuseness of one-body
potentials is changed due to the occupation, as we approach the drip lines.  
In the following the variation of shell structure is studied for 
potentials with a given diffuseness.
We note that in 
usual shell model calculations harmonic oscillator wave functions are used 
and single-particle energies are input parameters.  In order to give
appropriate 
input parameters for nuclei away from the $\beta$ stability line, 
it is important to know characteristic features of  both the variation of
kinetic energy and that of the strength of spin-orbit potential when
approaching the drip lines.
In ref. \cite{KS93} the isospin-dependence of kinetic energies in light
neutron-rich nuclei is suggested to be used in shell-model calculations.  We
will later comment on this suggestion.
In the present work we limit ourselves to the study of lighter nuclei, since
at the moment experimental data on nuclei close to the neutron drip line  
may be obtained only for lighter nuclei.

\begin{figure}[t]
\centering
\leavevmode
\epsfysize=8cm
\epsffile{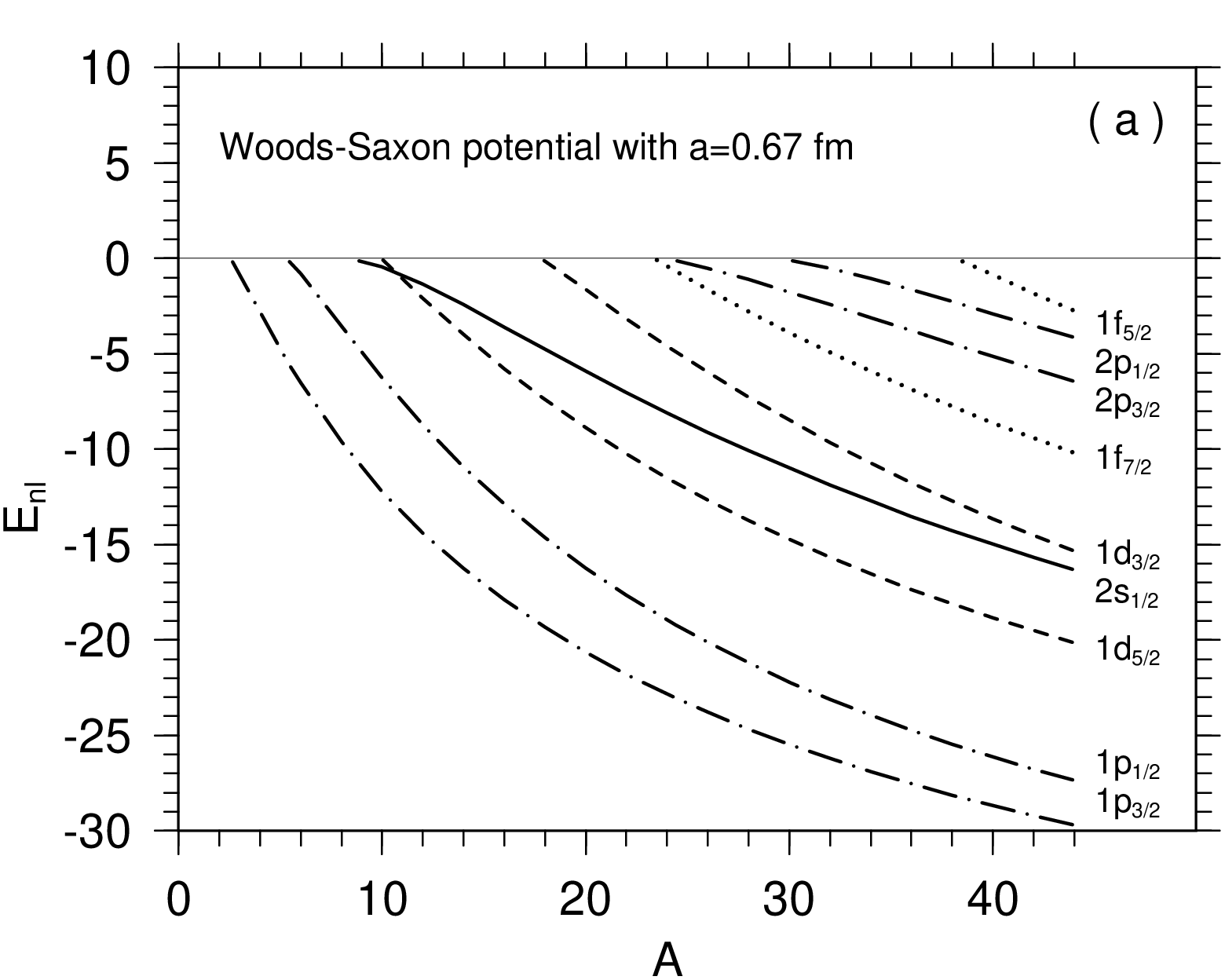}
\bigskip

\centering
\leavevmode
\epsfysize=8cm
\epsffile{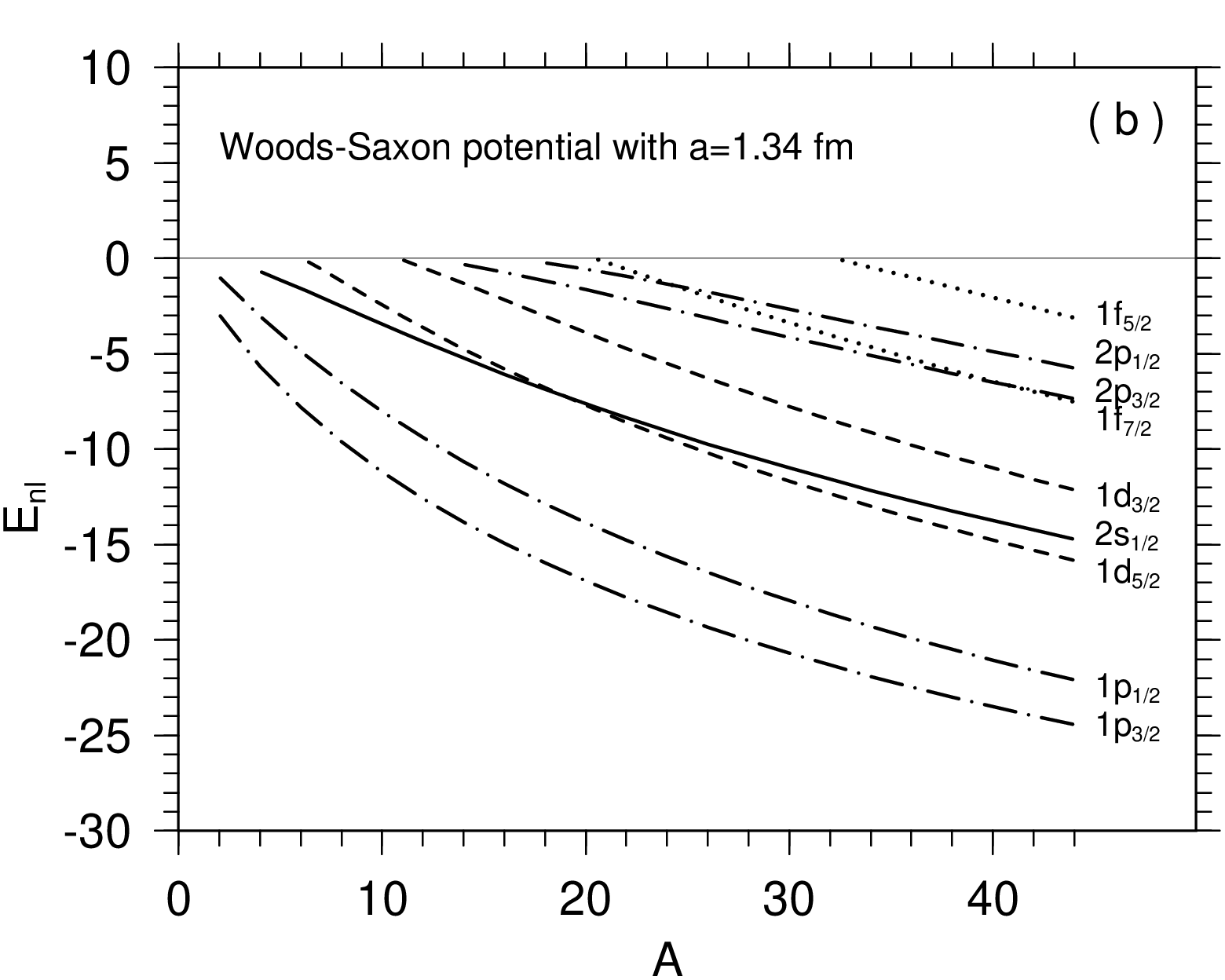}
\bigskip

\caption{Energies of neutron orbits : 
(a) for the Woods-Saxon potential with parameters, (\ref{eq:aws2}), (\ref{eq:aws3}),
 (\ref{eq:aws4}) and N=Z ; 
(b) the same as (a) except for $a$=1.34 fm.}
\label{fig1}
\end{figure}

\indent
In Fig. \ref{fig1}a we plot one-particle energy eigenvalues for neutrons in the
Woods-Saxon potential
\begin{equation}
U_{ws}(r)\: = \: - \, \frac{U_0}{1+exp(\frac{r-R}{a})} 
\label{eq:aws}
\end{equation}
together with the spin-orbit potential
\begin{equation}
U_{so} \: = \:  - \, v \, \left(\frac{\lambda_c}{2}\right)^2 \, \frac{1}{r} \, 
\frac{dU_{ws}(r)}{dr} \, (\vec{\ell} \cdot \vec{s}) 
\label{eq:aws1}
\end{equation}
choosing the 1p, 2s, 1d, 2p and 1f orbitals.  In the expression (\ref{eq:aws1})
$\lambda_c$ expresses the nucleon Compton wave-length.  Used parameters are 
\begin{eqnarray}
R & = & r_{0}A^{1/3} \qquad \mbox{with} \quad r_{0}=1.27 \quad \mbox{fm}
\nonumber \\
a & = & 0.67 \quad \mbox{fm} 
\label{eq:aws2}
\end{eqnarray}
and N=Z in the depth of the Woods-Saxon potential 
\begin{equation}
U_{0} \: = \:  \left(50 \, - \, 32\, \frac{N-Z}{A} \right) \quad \mbox{MeV}
\label{eq:aws3}
\end{equation}
and in the strength of the spin-orbit potential \cite{BM69}, 
\begin{equation}
v \: = \:  \left(32 \, - \, 20\, \frac{N-Z}{A} \right) 
\label{eq:aws4}
\end{equation}
The energies of neutron orbitals calculated using almost the same parameters
as in Fig. \ref{fig1}a are found for A$>$20 in p.239 of ref. \cite{BM69}. 
Due to the presence of less-bound neutrons the diffuseness of the potential 
may become larger as we approach neutron drip lines.  However, just using a
larger diffuseness may not always correspond to the realistic situation.   
This is partly because the major part
of the total potential comes from the core particles, which are well bound and 
are the majority.  Furthermore, it is 
partly because the proton core, which makes a more important contribution to 
the neutron potential than the neutron core, 
is more deeply bound in nuclei closer to neutron drip lines.
As a numerical example with a larger diffuseness, in Fig. \ref{fig1}b 
we show the one-particle energy eigenvalues for the Woods-Saxon potential with
the same parameters as those used 
in Fig. \ref{fig1}a except for doubling the diffuseness, $a$=1.34 fm.

\indent
In Figs. \ref{fig1}a and \ref{fig1}b we observe : (a) the A-dependence of the eigenvalues 
$E_{n\ell}$ in the energy region of 
$E_{n\ell} \! \simgeq \! -$15 MeV
depends very much on the orbital angular momentum ; (b) 
as is seen, for example, from the crossing of the eigenvalues of 1d$_{5/2}$ and 
2s$_{1/2}$ orbitals
at $E_{n\ell}$=$-$0.71 MeV in Fig. \ref{fig1}a and 
$E_{n\ell}$=$-$7.21 MeV in Fig. \ref{fig1}b, 
it is known that the slope of eigenvalues for lower $\ell$ orbitals becomes
relatively much less,  
as the eigenvalues approach zero ; (c) The spin-orbit splitting with a
given ($n,\ell$) may
generally be expected to become larger ($\sim$A$^{-1/3}$) 
as the mass number decreases.  In contrast, 
the spin-orbit splitting between 1p$_{1/2}$ and 1p$_{3/2}$ (or 
2p$_{1/2}$ and 2p$_{3/2}$) does not increase
but it rather decreases when $E_{n\ell}$ goes up from $-$5 to 0 MeV. In Fig. \ref{fig1}b
the decrease is observed also for the spin-orbit splitting of 1d orbitals; 
(d) For the potential with a larger diffuseness the features mentioned above, 
(a)-(c), are more pronounced.
We note that the prominent 
difference of one-particle shell structure,
such as the level order and the energy gap, between Figs. \ref{fig1}a and \ref{fig1}b is 
observed especially for $E_{n\ell}$$\simgeq$$-$10 MeV.

\indent
In ref. \cite{HZ98} 
we studied the coupling of particles near threshold to shape oscillations and
showed the evaluated matrix elements of
\begin{equation}
V_{pv} (r)\: = \: R \, \frac{dU(r)}{dr} \qquad 
\label{eq:a1}
\end{equation}
where $U(r)$ expresses the one-particle potential while $R$
represents the nuclear radius.
Noting that 
the kinetic energy of the spherical system in the absence of spin-orbit
potential is equal to the expectation value of 
\begin{equation}
\frac{r}{2} \, \frac{dU(r)}{dr} 
\label{eq:a2}
\end{equation}
due to the virial theorem, 
for the square-well potential the expectation value 
of (\ref{eq:a1}) is equal to 
twice the kinetic energy. 
Furthermore, the expectation value of 
\begin{equation}
\frac{1}{r} \frac{dU(r)}{dr} \qquad 
\label{eq:a3}
\end{equation}
which is the radial dependence of the spin-orbit potential, 
is equal to 1/$R^2$ times that of (\ref{eq:a1}) in the case of
the square-well potential.  Thus, for the square-well potential Fig. 3a of
ref. \cite{HZ98} already contains all information on both the 
kinetic energy and the spin-orbit splitting.  
In the present work we investigate those two quantities for Woods-Saxon
potentials compared with the square-well potential,
studying the dependence on energy especially 
for E$_{n\ell}$=$-$10$\rightarrow$0 MeV 
and on angular momenta.
Though it can be disputed  
whether the radial derivative of the density or that of the
potential should be used 
for the radial dependence of the spin-orbit potential, for simplicity we
use the form in (\ref{eq:a3}).

\indent
In sect.2 we present numerical results and discussions, while conclusions 
are given in sect.3.

\section{Numerical results and discussions}
\begin{figure}[t]
\centering
\leavevmode
\epsfysize=6.5cm
\epsffile{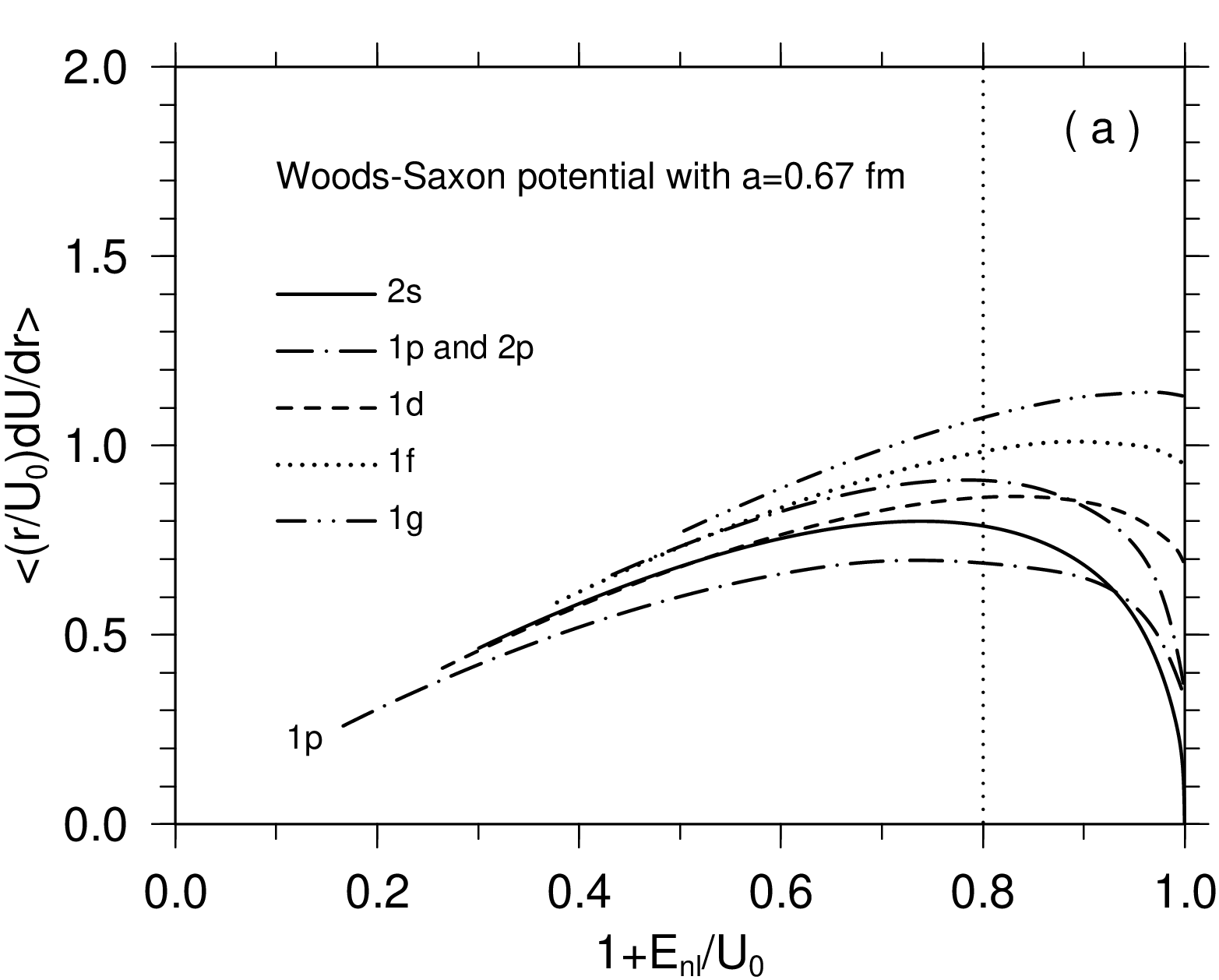}

\centering
\leavevmode
\epsfysize=6.5cm
\epsffile{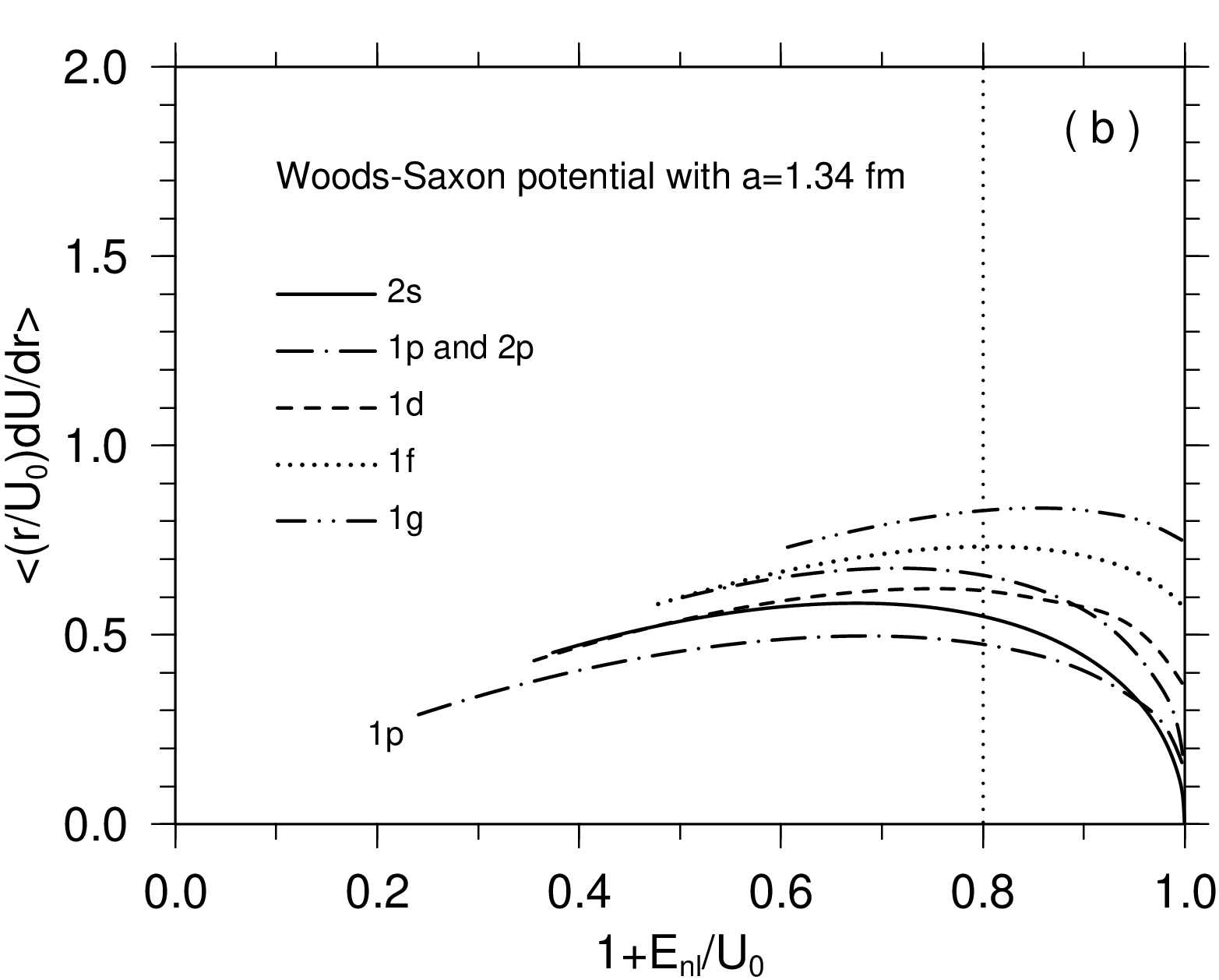}

\centering
\leavevmode
\epsfysize=6.5cm
\epsffile{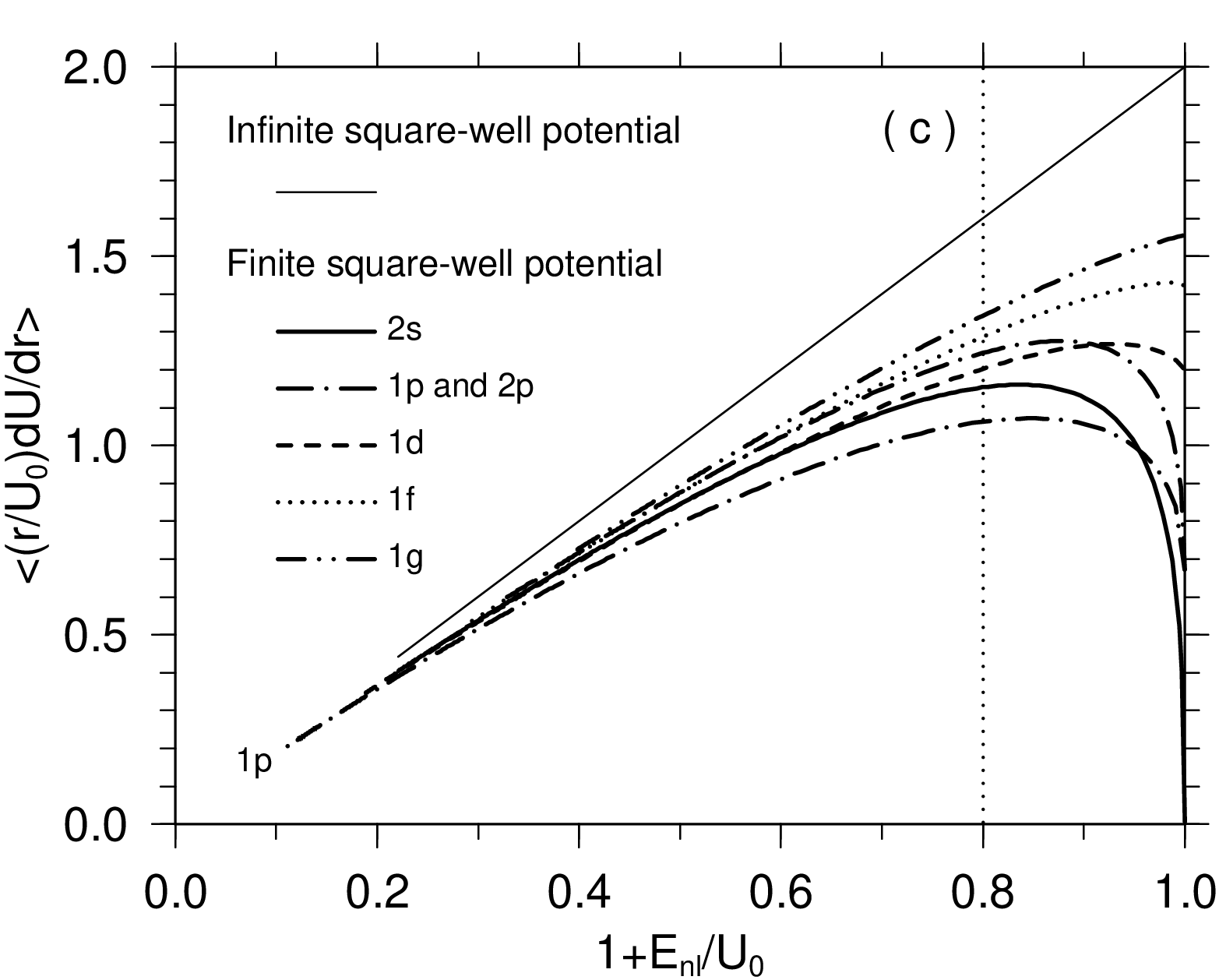}
\bigskip
\caption{The expectation values of the operator (\ref{eq:b1}), which is proportional to
kinetic energy of respective one-particle orbitals, as a function of 
(1+(E$_{n\ell}$/U$_0$)) where E$_{n\ell}$ expresses the one-particle energy 
eigenvalue of the orbital ($n , \ell$).
For U$_0$=50 MeV (1+(E$_{n\ell}$/U$_0$))=0.8 and 1.0 
mean E$_{n\ell}$=$-$10 and 0
MeV, respectively.
(a) for the Woods-Saxon potential with parameters,
(\ref{eq:aws2}), (\ref{eq:aws3}), $v$=0 and N=Z ; (b) the same as (a)
except for $a$=1.34 fm ; (c) for the square-well potential (\ref{eq:b3}).
By the thin straight line in Fig. \ref{fig2}c the expectation value of the operator 
(\ref{eq:b1}) is
exhibited for the infinite square-well potential.
The thin line is the same for all orbitals with  
various ($n , \ell$) values.
See the text for details.}
\label{fig2}
\end{figure}

\begin{figure}
\centering
\leavevmode
\epsfysize=6.5cm
\epsffile{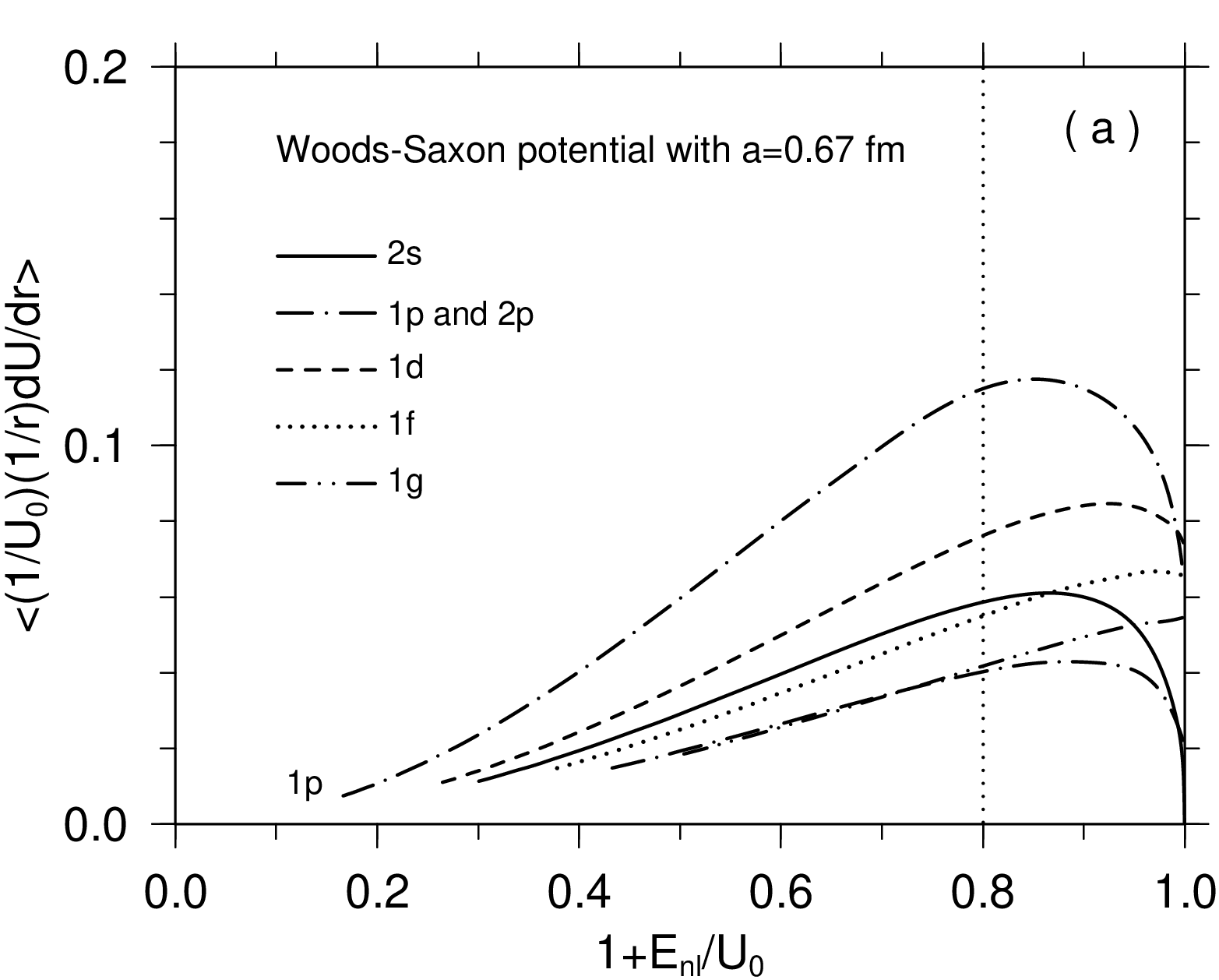}

\centering
\leavevmode
\epsfysize=6.5cm
\epsffile{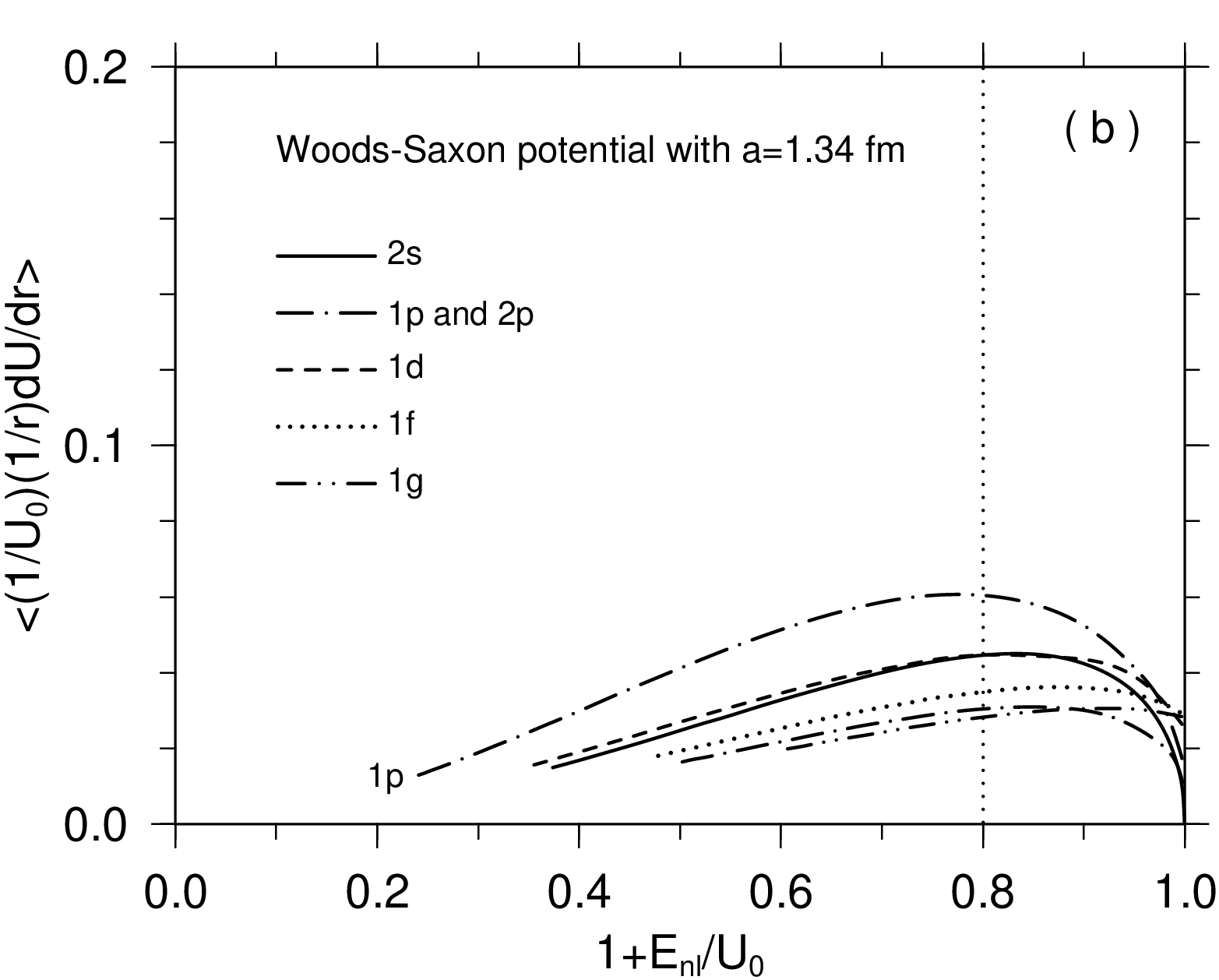}

\centering
\leavevmode
\epsfysize=6.5cm
\epsffile{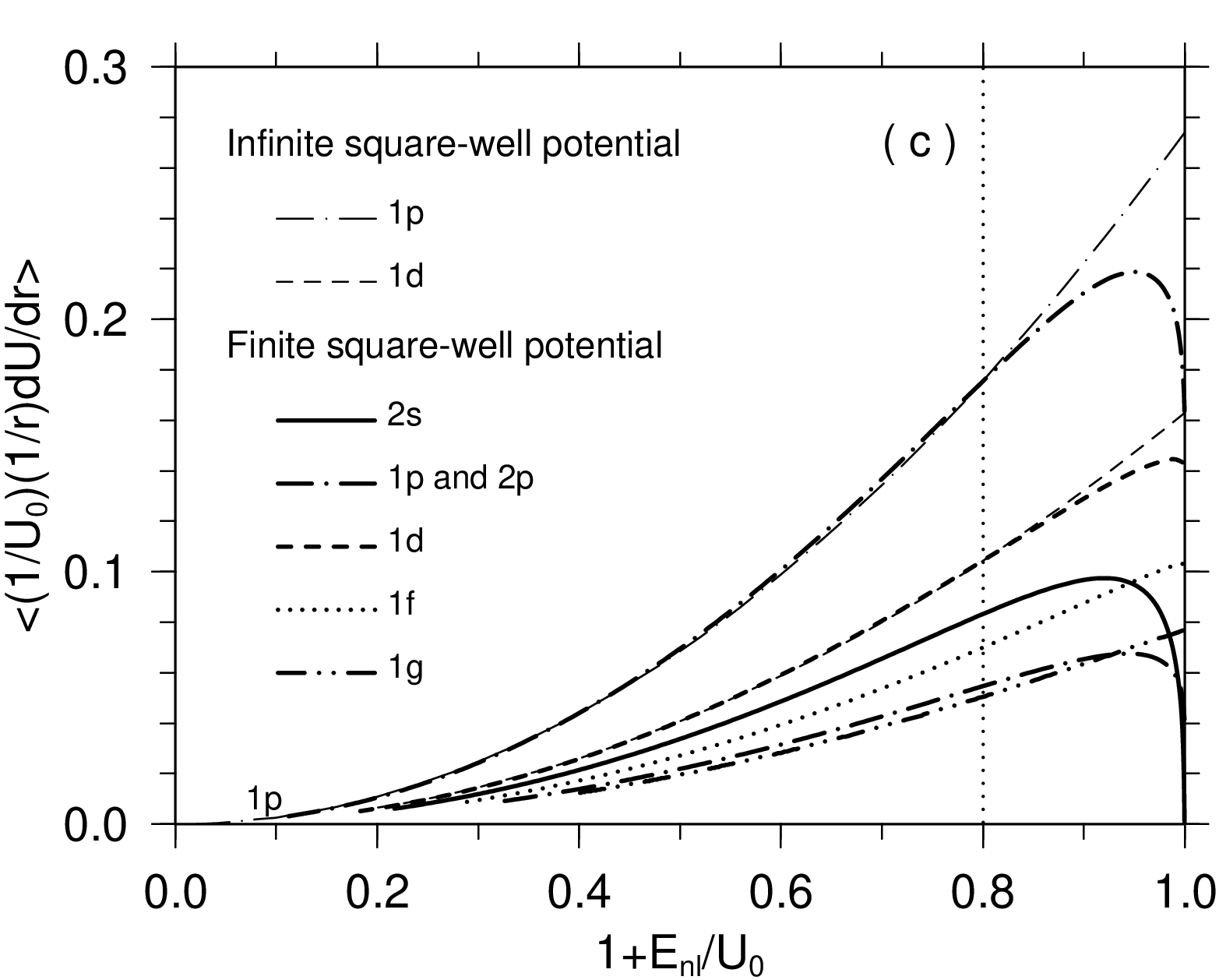}
\bigskip
\caption{The expectation values of the operator (\ref{eq:b2}), which in the perturbation
treatment of spin-orbit potential is proportional to
the spin-orbit splitting of the $\ell$ orbital 
when multiplied by (2$\ell$+1), as a function of 
(1+(E$_{n\ell}$/U$_0$)). 
For U$_0$=50 MeV (1+(E$_{n\ell}$/U$_0$))=0.8 and 1.0 
mean E$_{n\ell}$=$-$10 and 0
MeV, respectively.
(a) for the Woods-Saxon potential with parameters,
(\ref{eq:aws2}), (\ref{eq:aws3}), $v$=0 and N=Z ; (b) the same as (a)
except for $a$=1.34 fm ; (c) for the square-well potential (\ref{eq:b3}) with 
U$_0$=50 MeV.
By the thin curves in Fig. \ref{fig3}c the expectation value 
in the infinite square-well potential 
is shown for the 1p and 1d orbitals.
See the text for details.}
\label{fig3}
\end{figure}

\indent
In the present section we do not include 
the spin-orbit potential in the one-body potential, 
for simplicity.
We evaluate the expectation value of the operators,
\begin{equation}
\frac{r}{U_0} \, \frac{dU(r)}{dr} 
\label{eq:b1}
\end{equation}
and 
\begin{equation}
\frac{1}{r \, U_0} \, \frac{dU(r)}{dr}  
\label{eq:b2}
\end{equation}
for neutron orbitals in Woods-Saxon potentials.
The expectation values of (\ref{eq:b1}) is equal to those
plotted in Fig. \ref{fig3}a of \cite{HZ98} in the case of the square-well potential.
Note that the quantity on the x-axis in Fig. \ref{fig3}a of ref. 
\cite{HZ98} for the square-well potential has the same meaning
as that on the x-axis in Fig. \ref{fig2} and \ref{fig3} of the present paper, though the notation
used for the square-well potential in ref. \cite{HZ98} was different.
The expectation value 
of (\ref{eq:b1}) is proportional to the kinetic energy, while 
that of (\ref{eq:b2}) to the spin-orbit splitting for a given orbital
angular momentum $\ell$ treating the spin-orbit potential by perturbation.
We examine those expectation values for a given 
($n$,$\ell$) value as a function
of E$_{n\ell}$/$U_0$ where E$_{n\ell}$ expresses one-particle energy 
eigenvalues.

\indent
In Figs. \ref{fig2}a and \ref{fig2}b the expectation value of the operator 
(\ref{eq:b1}) is plotted for the Woods-Saxon potential with the diffuseness 
$a$=0.67 and 1.34 fm, respectively, while in Fig. \ref{fig2}c
we show the same quantity for the square-well potential, 
\begin{equation}
U(r) \:= \: \left\{ \begin{array}{ll}
-U_{0} 
& \qquad \mbox{for} \quad r < R \\
\\
0
& \qquad \mbox{for} \quad r \geq R 
\end{array}
\right.
\label{eq:b3}
\end{equation}
a part of which was
taken from Fig. \ref{fig3}a of ref. \cite{HZ98}.
In the limit of infinite square-well potential
(U$_{0}$$\rightarrow$$\infty$), the kinetic energy is equal to U$_{0}$+E$_{n
\ell}$ and, thus, the expectation value of (\ref{eq:b1}) multiplied by U$_{0}$ 
is given by
2(U$_0$+E$_{n\ell}$) for all orbitals with various ($n , \ell$) values. 
The expectation value for the infinite
square-well potential is shown in Fig. \ref{fig2}c by the thin straight line.
Since the relation between the quantities in the x- and y-axis multiplied by
U$_{0}$ is expressed by a straight line, dividing both quantities by any given
``U$_{0}$'' does not change the straight line.  Thus, we plot the line for  
the infinite square-well potential also in Fig. \ref{fig2}c.
Note that one-particle
wave-functions cannot extend to outside of the potential with 
U$_{0}$$\rightarrow$$\infty$ and, thus, the reduction in the increasing rate 
of kinetic energy does not occur as 
(1+(E$_{n \ell}$/U$_{0}$)) increases.

\indent
It should be noted that along the curve with a given ($n$,$\ell$) 
the mass number monotonically decreases as (1+(E$_{n\ell}$/U$_0$)) increases.
For a very large value of U$_{0}$, one may approximately 
write (see eq. (\ref{eq:b4}))
\begin{equation}
U_{0} \, + \, E_{n \ell} \: = \: \frac{c_{n \ell}}{R^2}
\label{eq:b3a}
\end{equation}
where c$_{n \ell}$$>$0, U$_{0}$$>$0 and $E_{n \ell}$$<$0, while R expresses the
radius of the system. 
Thus, in order to obtain larger eigenvalues $E_{n \ell}$, namely 
larger values of (1+(E$_{n\ell}$/U$_0$)), keeping a fixed
value of U$_{0}$ , one must decrease values of R (or the mass number A).

\indent
The vertical dotted line of Fig. 2 at 1+(E$_{n\ell}$/U$_0$)=0.8, namely
E$_{n \ell}$=$-$10 MeV for U$_0$=50 MeV, indicates the
approximate Fermi level in $\beta$ stable nuclei.
When the eigenvalue E$_{n\ell}$$\rightarrow$0 exists for the potential with
a finite range, 
the expectation value of (\ref{eq:b1})
for ($n,\ell$=0) orbitals in the finite-well
potential approaches zero, 
since due to the absence of centrifugal barrier the $\ell$=0 neutron wave
functions can extend up to infinity.
For the square-well potential the expectation value of (\ref{eq:b1}) at 
E$_{n\ell}$=0 is independent of $n$ and 
is equal to 2/3, 6/5 and 10/7, for the $\ell$=1, 2 and 3 orbitals, respectively
\cite{HZ98}. 
In contrast, the expectation value for Woods-Saxon 
potentials depends on radial node
$n$ and is almost a factor two (four) smaller 
for the potential in Fig. \ref{fig2}a (\ref{fig2}b).
It is also noted that for more diffuse potentials 
the absolute magnitude of
kinetic energy of a given ($n,\ell$) orbital is considerably smaller for  
almost all values of $E_{n\ell}$.

\indent
The effect of the diffuse surface of the potential on kinetic energy is seen
from the comparison of the curves in Figs. \ref{fig2}a and \ref{fig2}b with those in Fig. \ref{fig2}c
for the finite and infinite square-well potential. 
The considerable deviation
from the straight line occurs first at smaller values of $\mid \! E_{n\ell}
\! \mid$ for less diffuse potential and for larger $\ell$ orbitals.  In other
words,  
the kinetic energy starts to decrease already 
at relatively larger binding energy for more diffuse potentials.
This is because for more diffuse potentials the wave functions can more easily
extend to the outside of the potentials.  The extension is easier for the
orbitals with smaller
orbital angular momenta because of the lower centrifugal barriers.
As seen in Fig. \ref{fig2}c, for the square-well potential the kinetic energy starts to
decrease first at very small binding energies, even for $\ell$=0 orbitals.

\indent
The deviation of kinetic energy of ($n,\ell$) orbitals 
from the monotonic increase 
as E$_{n \ell}$ increases from $-$10 to 0 MeV
depends somewhat on the radial node $n$.  The larger the quantum
number $n$, the larger the variation of the kinetic energy   
for E$_{n \ell}$=$-$10 $\rightarrow$ 0 MeV. 
This is understood because at 
(1+E$_{n\ell}$/U$_0$)=0.8, namely E$_{n\ell}$=$-$10 MeV, 
the kinetic energy of orbitals with larger $n$ is
larger than that with smaller $n$, while at E$_{n \ell}$=0 the
kinetic energy of orbitals with a given $\ell$ is almost independent of $n$. 
The latter is independent of $n$ for the square-well potential. 

\indent
In ref. \cite{KS93} the isospin dependence of kinetic energies in very light
neutron-rich nuclei is suggested, in which the isospin dependence for the
2s$_{1/2}$ and 1d$_{3/2}$ state is taken to be the same as that for 
the 1d$_{5/2}$ state.
The origin of the isospin-dependence 
is certainly traced back to 
the decreasing binding energy of least bound neutrons as well as 
the possible larger
surface diffuseness of the potential, when the
isospin increases in those light neutron-rich nuclei. 
However, from our study shown in Fig. \ref{fig2} we note that the 
binding-energy dependence of kinetic energy for the 2s and 1d states are very
different for a given potential, 
especially 
for E$_{n \ell}$=$-$10 $\rightarrow$ 0 MeV.
Observe the
difference of the behaviour between the solid and dashed curve in Fig. \ref{fig2} for 
(1+E$_{n\ell}$/U$_0$)=0.8$\rightarrow$1.0 .

\indent
In Figs. \ref{fig3}a and \ref{fig3}b the expectation value of the operator 
(\ref{eq:b2}) is plotted for the Woods-Saxon potential with the diffuseness 
$a$=0.67 and 1.34 fm, respectively, while in Fig. \ref{fig3}c
we show the same quantity for the square-well potential.
For the
square-well potential the expectation value of (\ref{eq:b2}) depends on some
extra parameter to specify, either U$_0$ or $R$, other than the only parameters
unique in the model, U$_0$$R^2$ and 
E$_{n\ell}$$R^2$ \cite{RJM92}.  We have used U$_{0}$=50 MeV. 

\indent
The thin curves in Fig. \ref{fig3}c are estimated for the infinite square-well
potential.  For the (U$_{0}$$\rightarrow$$\infty$) square-well potential
the kinetic energy is equal to 
\begin{equation}
U_{0} \, + \, E_{n \ell} \: = \: \frac{\hbar^2 \, x_{n \ell}^2}{2 \, M \, R^2}
\label{eq:b4}
\end{equation}
where $x_{n \ell}$ expresses the $n$-th zero point of the spherical Bessel
function $j_{\ell}$(z). Thus, the expectation value of the operator in 
(\ref{eq:b2}) multiplied by U$_{0}$ is written as 
\begin{equation}
\frac{2(U_{0} \, + \, E_{n \ell})}{R^2} \: = \: 
\frac{4 \, M \, (U_{0} \, + \, E_{n \ell})^2}{\hbar^2 \, x_{n \ell}^2} \: .
\label{eq:b5}
\end{equation}
Variables in the x- and y-axis in Fig. 3c are those 
in (\ref{eq:b4}) and (\ref{eq:b5}) divided by ``U$_{0}$'', however, the
division by ``U$_{0}$''=$\infty$ does not make sense.
Since the relation between the quantities 
in (\ref{eq:b4}) and (\ref{eq:b5})
is not linear, the shape of the curve depends on the ``U$_{0}$'' value used for
evaluating variables in the x- and y-axis.
Since our purpose is to compare the result of finite potentials with that of
the infinite potential, we 
choose this ``U$_{0}$'' value so that at
(1+(E$_{n \ell}$/U$_{0}$))=0.8 the thin curve coincides with the corresponding
curve for the finite square-well potential.
It is seen that only in the region
(1+(E$_{n \ell}$/U$_{0}$))$\simgeq$0.8 
the curve for the finite square-well potential differs
appreciably from that for the infinite square-well potential.
 
\indent
The plotted quantity in Fig. \ref{fig3} times (2$\ell$+1) is proportional to 
the spin-orbit splitting.
For a given E$_{n \ell}$ value the spin-orbit splitting 
for the 1p orbital is much larger 
than that for the 2p orbital, since the former belongs to lighter
nuclei.  For the purpose of comparison of expectation values of (\ref{eq:b2}), 
we plotted calculated values also for 2s orbitals, though
no spin-orbit potential acts on s-orbitals.

\indent
The diffuse surface effect of potentials on spin-orbit splitting is seen from
the comparison of curves in Figs. 3a and 3b 
with those in Fig. 3c for the finite and infinite square-well
potential.
Due to the different r-dependence, the effect of the diffuse surface on kinetic
energy starts to appear already at deeper bound orbitals compared with 
spin-orbit splitting.
As already exhibited in Fig. \ref{fig1}b, for the Woods-Saxon potential with 
$a$=1.34 fm the
spin-orbit splitting of 1d orbitals decreases also appreciably 
as E$_{n\ell}$ changes from
$-$10 to 0 MeV.  The decrease corresponds to the variation of the dashed
curve in Fig. \ref{fig3}b for (1+E$_{n\ell}$U$_0$)=0.8$\rightarrow$1.0 .

\indent
The variation of spin-orbit splitting of orbitals with a given $\ell$ 
for E$_{n\ell}$=$-$10$\rightarrow$0 MeV
depends moderately on the radial node $n$.   
We also note that the comparison of spin-orbit splittings of
the orbitals with a given $\ell$ but with different $n$ means the comparison
made for nuclei with different mass number.

\indent
In order to simulate the spin-orbit potential in which the radial derivative of
the density is used instead of that of the potential, we have also made the
numerical calculation using $r_0$=(0.9)(1.27)=1.14 fm in U(r) of (\ref{eq:b2}) 
while keeping
one-particle wave-functions unchanged.  The dependence of the expectation value
of (\ref{eq:b2}) on either $\ell$ or $n$ 
for E$_{n\ell}$=$-$10$\rightarrow$0 MeV 
is found to be almost the same as what is described above.

\section{Conclusions}
\indent
In order to study the variation of one-particle shell-structure due to 
the surface diffuseness of nuclear potential, the radial matrix-elements of
both kinetic energy and spin-orbit potential of Woods-Saxon potentials 
are estimated and compared with those
of the infinite and finite square-well potential. 
We are especially interested in the energy region where 
binding energies of neutrons
decrease from 10 to 0 MeV, having nuclei close to neutron drip lines in mind.
In the infinite square-well potential neither kinetic energy nor spin-orbit
splitting ever decreases as a function of E$_{n \ell}$.  For the finite
square-well potential with U$_{0}$=50 MeV a decrease is observed for 
E$_{n \ell}$=$-$10$\rightarrow$0 MeV, depending on orbitals.
For orbitals with smaller $\ell$ values both quantities decrease
more strongly as binding energies approach zero.
The decrease is stronger for more diffuse potentials.
Due to the different r-dependence, the effect of the diffuse surface on kinetic
energy starts to appear already at deeper bound orbitals compared with 
spin-orbit splitting.

\indent
In traditional shell model calculations carried out for 
neutron drip line nuclei, 
it is important to take into account 
the variation of both kinetic energy and spin-orbit
splitting in the input one-particle energies, 
which is studied in the present work.   
Not only the binding energy dependence of those quantities 
but also the $\ell$ dependence is 
very important.\\

\noindent
{\large{\bf Acknowledgements}}\\
X.Z.Z. acknowledges the financial support provided by 
Crafoordska Stiftelsen, while S.V.L. is grateful to The Royal Swedish Academy
of Sciences for the financial support under the contract for cooperation
between Sweden and the former Soviet Union.

\newpage

\newpage

\newpage

\end{document}